\documentclass[useAMS,usenatbib]{mn2e}
\usepackage{epsfig}
\usepackage{url}

\usepackage[T1]{fontenc}
\usepackage{aecompl}
\usepackage{color}

\newcommand{\msun}{M_{\odot}}

\definecolor{simonacolor}{cmyk}{0, 0.9808, 0.4429, 0.1412}
\definecolor{simonacolor2}{cmyk}{0.6, 0.0, 0.99, 0.21}

\definecolor{richardcolor}{cmyk}{0., 0.5, 1, 0}
\definecolor{richardcolor2}{cmyk}{0.6, 1.0, 0.99, 0.21}

\title[Stellar Mass Function]{The Massive End of the Stellar Mass Function}

\author[R. D'Souza, S.Vegetti, G. Kauffmann]
{Richard D'Souza$^{1}$\thanks{E-mail address: rdsouza@mpa-garching.mpg.de
(RDS)}, Simona Vegetti$^{1}$,  Guinevere Kauffmann$^{1}$\ \\
$^{1}$Max Plank Institute for Astrophysics, Munich, Germany\\
}

\begin{document}

\date{Accepted 1988 December 15. Received 1988 December 14; in original form 1988 October 11}

\pagerange{\pageref{firstpage}--\pageref{lastpage}} \pubyear{2002}

\maketitle

\label{firstpage}

\begin{abstract}
  We derive average flux corrections to the \texttt{Model} magnitudes of the Sloan Digital Sky Survey (SDSS)
  galaxies by stacking together mosaics of similar galaxies in bins of stellar mass and concentration.
  Extra flux is detected in the outer low surface brightness part of the galaxies, 
  leading to  corrections ranging from 0.05 to 0.32 mag for the highest stellar mass galaxies.
  We apply these corrections to the MPA-JHU (Max-Planck Institute for Astrophysics - John Hopkins University)
  stellar masses for  a complete sample of half a million galaxies from the SDSS survey to derive
  a corrected galaxy stellar mass function at $z=0.1$ in the stellar mass range  $9.5<\log(M_\ast/M_\odot)<12.0$. We find that the
  flux corrections and the use of the MPA-JHU stellar masses have a significant impact on the massive end of
  the stellar mass function, making the slope significantly shallower than that estimated by Li \& White (2009),
  but steeper than derived by Bernardi et al. (2013). This corresponds to a mean comoving stellar mass density
  of galaxies with stellar masses $\log(M_\ast/M_\odot) \ge 11.0$ that is a factor of 3.36 larger than the
  estimate by Li \& White (2009), but is 43\% smaller than reported by Bernardi et al. (2013).

\end{abstract}


\section{Introduction}
The stellar mass function of galaxies is a  basic probe of galaxy formation and evolution enabled by large redshift surveys.
In recent years, major advances have been made by large redshift surveys, such as the 2dF Galaxy Redshift Survey and the
Sloan Digital Sky Survey (SDSS), in estimating the stellar mass function in the low-redshift Universe \citep{Cole2001, Bell, Blanton2003}.
For example, \cite{Li2009}  have used a uniform sample of almost half a million galaxies from SDSS DR7 to derive the
stellar mass function at $z=0.1$. This has been complemented by the effort of the Galaxy and Mass Assembly Survey
\citep[GAMA][]{Baldry}, which has accurately constrained the faint end slope of the stellar mass function
down to stellar masses $\sim 10^{8} \msun$.

The calculation of the stellar mass function hinges on the proper determination of the stellar mass of a galaxy,
which in turn depends critically on the estimation of its total flux in a given pass-band. Systematic differences
in the estimation of the stellar mass of a galaxy may arise from different choices of the initial mass function 
(IMF) and the stellar mass-to-light ratio (M/L), as well as from different estimations of the galaxy total flux. Determining the flux
accurately for a large number of galaxies in an all-sky survey is a challenging task. In particular, quantifying
the flux in the outer low surface brightness (LSB) regions of a galaxy has proven to be difficult and is still
subject of much debate \citep{Bernardi,Simard}. These uncertainties mean that the slope at the massive end of the
mass function is not very well determined. This has significant implications for several 
astrophysical problems, including halo occupation models, the mean baryon fraction in the Universe,  X-ray and
Sunavey-Zeldovich studies of high mass galaxies, and understanding the evolution of massive galaxies to
high redshifts.

Different approaches have been employed by SDSS in its photometric pipeline (\texttt{PHOTO}) to estimate
the total flux of a galaxy. In addition to SDSS \texttt{Petrosian} magnitudes,  two dimensional models
(e.g. exponential or de Vaucouleurs) have been used to model the surface brightness distribution of
galaxies (SDSS \texttt{Model} magnitudes).  Further improvement has been provided by SDSS \texttt{cModel}
magnitudes, for which fluxes are estimated as a linear combination of an exponential  and a de Vaucouleurs model.
In recent years, several studies have tried to fit S\'{e}rsic and multi-component models to the surface
brightness distribution \citep{Simard, Lackner, Bernardi}.

Each of these approaches provides a progressively better estimate of the total flux of a galaxy, but they
all suffer from the same intrinsic drawback, namely that the models are
fits to the central, high signal-to-noise ratio (SNR) regions of the galaxy 
 and assumptions are required about the outer lower SNR
(beyond $\mu_r \sim 27 \, \mathrm{mag\,arcsec}^{-2}$) part of the galaxy profile. Additionally, the total
flux estimated through model fitting can be biased in a number of ways.

The biggest source of systematic bias in the flux determination is related to the estimation of the sky background,
especially for large nearby objects or those located in dense environments \citep{linden2007, Bernardi2007}.
In principle, this can be overcome by considering extremely large fields of view. For example, considerable
progress has been achieved by \cite{Blanton2011} by fitting the masked sky background for each SDSS scan with 
a smooth continuous function. 

However, even with improvements to the sky background algorithm, one is still limited by the depth
of the survey. The relatively short exposure time of SDSS (53.9 secs) limits the accuracy of the background
determination and subtraction. This in turn limits ones ability to distinguish between the flux of the
outer stellar halo and the sky background, leading to an over- or under-estimation of the total flux of a
galaxy. In particular,  multi-component model fitting of the main galaxy
can lead to biased results. This may explain why recent attempts to trace the
low SNR LSB part of a galaxy through fitting multi-component models to single SDSS photometric images have 
yielded divergent results \citep{Simard, Bernardi, Meert2015}.

Other sources of systematic error in determining the flux of a particular object are the procedures employed 
for deblending and masking, as well as the radial extent of the models used for the surface brightness fitting.
Finally, in addition to photometry, several other effects have a considerable impact on the massive end 
of the stellar mass function, such as evolutionary corrections and 
\emph{fiber collisions} (i.e. the fraction of galaxies not targeted for spectroscopy due to the fact that
fibres cannot be positioned closer together than 55 arcseconds on the SDSS plug plates).

An alternative but viable approach to fitting models to individual images of galaxies, is to stack images of similar 
galaxies to quantify the average total amount of extra light in the outer parts \citep{Tal, DSouza}. By stacking galaxies as a 
function of their stellar mass and galaxy-type, \cite{DSouza} have reached a depth of $\mu_r \sim 32 \, \mathrm{mag\,arcsec}^{-2}$. 
The increased depth of galaxy stacks helps to reliably constrain the total amount of light especially in the 
LSB component. In addition, the background for stacked galaxies can be determined more accurately. This then provides a
direct handle on the corrections to the \texttt{Model} magnitudes as a function of the stellar mass and galaxy type. 

In this paper, we attempt to derive flux corrections to the \texttt{Model} magnitudes and re-derive the galaxy
stellar mass function at redshift $z=0.1$ using MPA-JHU (Max-Planck Institute for Astrophysics
  \& John Hopkins University) stellar masses \citep{Kauffmann03a} and  the sample of
\cite{Li2009}. We estimate corrections to the \texttt{Model} magnitudes by stacking volume-limited samples in bins of
stellar mass, concentration and model type. We also consider various effects that may systematically
bias the stellar mass function.

In Section \ref{sec:sample}, we define the samples used for deriving the corrections
as well as the full sample used to derive the stellar mass function.
In Section \ref{sec:corrections}, we derive the flux corrections to
the \texttt{Model} magnitudes. In Sections \ref{sec:massfunction} and \ref{sec:lumfunction},
we derive the galaxy stellar mass function and the luminosity function respectively. In Sections
\ref{sec:summary} and \ref{sec:discussion}, we summarise and discuss our results.
Throughout this paper, we assume a flat $\Lambda$CDM cosmology, $\Omega_{\mathrm{m}}=0.25$ and
$\Omega_{\mathrm{\Lambda}}=0.75$. We further assume a Hubble parameter $h=0.72$ for the calculation
of physical distance scales wherever necessary.

\section{Sample Selection}
\label{sec:sample}
\subsection{Sample for Calculating the Mass Function}
Following \cite{Li2009}, we select SDSS spectroscopic galaxies from the NYU-VAGC (New York University - Value Added Catalogue)
\footnote{Available at http://sdss.physics.nyu.edu/vagc/ .} catalogue \citep{Blanton2005b} with 
redshifts in the range $0.001\,\le\,z\le\,0.5$ and  \texttt{Petrosian} $r$-band magnitudes in the range 
$12\,\le\,m_{r Pet}\,\le\,17.6$.\footnote{We also include the three survey strips in the Southern Cap.} This gives us a total 
of 533442 galaxies, which are ideal for large scale structure studies. 
We further pruned the sample to 523476 galaxies by retaining only those galaxies with a
valid MPA-JHU stellar mass. We estimate the ``effective'' survey area to be  $6570\,deg^{2}$
(2.0084 steradians), by taking into account the incompletness and the masked-out regions
(due to bright stars) of the survey.

For the stellar mass function, we use the stellar masses provided in the DR7 
version of the MPA-JHU catalogue\footnote{Available at http://www.mpa-garching.mpg.de/SDSS/DR7/ },
which assumes a universal Chabrier  initial stellar mass function ( Chabrier 2003).

To derive the luminosity function, we use the r-band absolute \texttt{Model} magnitude ($M_{r^{0.1}}$),
corrected for evolution and K-corrected to its value at $z_{0}=0.1$ according to the following equation:
\begin{equation}
  \label{eq:mag}
  M = m - DM(z) - K(z;z_{0}) + Q_{e}(z-z_{0})\,,
\end{equation}
where $M$ is the absolute magnitude, $DM(z)$ is the distance modulus at redshift $z$, 
$m$ the apparent magnitude, $K(z;z_{0})$ is the $K$-corrections relative to a passband blue-shifted
by $z_{0}$ and the luminosity $e$-correction is parametrised linearly by $Q_{e}$. The $K$-corrections
were calculated using the code \texttt{Kcorrect} \texttt{v4\_3} \citep{Blanton2007}. In general, we
assume a uniform luminosity evolutionary correction of $Q_{r}=1.62$ as  derived by \citep{Blanton2003}.

\subsection{Sample for Determining the Flux Corrections}
\label{sec:sample_stacks}
To derive the corrections to the \texttt{Model} magnitudes, we stack volume-limited 
sub-samples of isolated galaxies defined from the parent sample in various ranges of stellar 
mass, concentration ($R90/R50$) and redshift (See Table \ref{tbl:samples}). 
In each sub-sample, galaxies that were better fit by an exponential (\texttt{Exp}) or a 
de Vaucouleurs (\texttt{deV}) model by the SDSS pipeline (defined by comparing the likelihood values 
of the model fits from the SDSS \texttt{PhotoObjAll} database) were stacked separately.

We select isolated galaxies by requiring that there are no brighter companions in the spectroscopic sample
within $R \le 1 \, \mathrm{Mpc}$ (where $R$ is the projected comoving separation) and $|\delta z| < 1000  \, \mathrm{km\, s}^{−-1}$.

\begin{figure}
  \begin{center}
    \includegraphics[width = 0.48\textwidth]{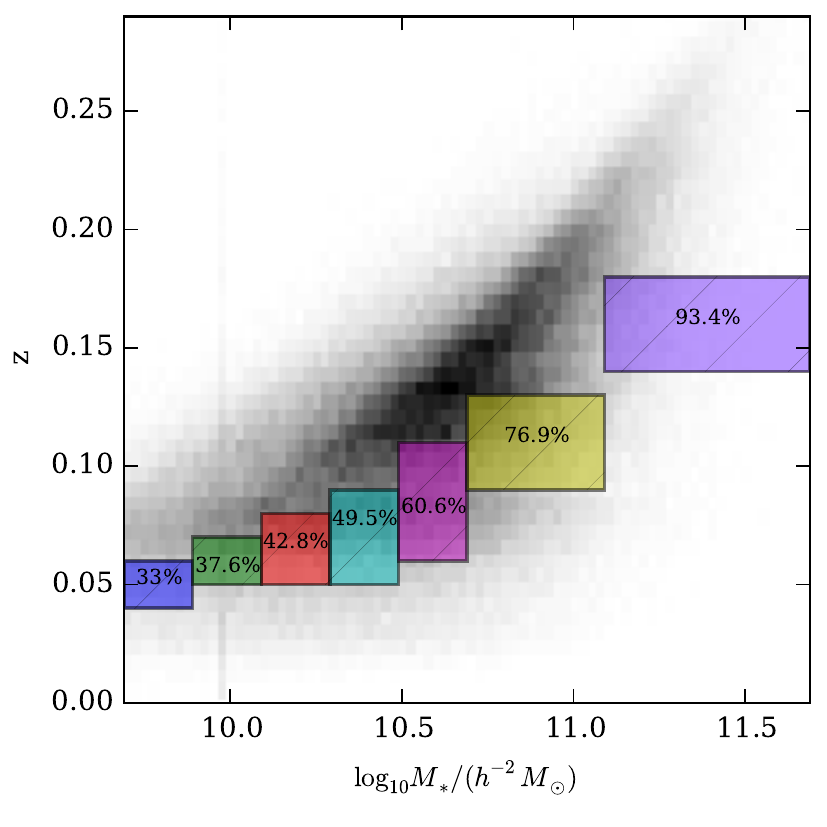}
  \end{center}
  \caption{Volume-limited samples used for stacking: Shaded contours show the distribution of
    the parent sample galaxies in the plane of stellar mass versus redshift. The seven coloured boxes
    indicate the redshift limits of the seven stellar mass sub-samples. These sub-samples are further
    divided by concentration (C) (not shown in the figure). 
    The numbers in the coloured boxes indicate the fraction of centrals in these
    volume limited stellar mass sub-samples.
    } 
 \label{fig:redshiftcuts} 
\end{figure}

In Figure \ref{fig:redshiftcuts}, the redshift limits of the various stellar mass sub-samples are shown
projected along the plane of stellar mass versus redshift of the parent NYU-VAGC sample. The fraction
of centrals in each stellar mass sub-sample are also shown.

\begin{table*}
  \caption{Volume-limited samples of isolated galaxies selected by stellar mass from the
    NYU-VAGC sample for the purpose of stacking}.
  \begin{center}
    \begin{tabular}{cccccc}\hline\hline
      Sample & Stellar mass & Concentration & Redshift & $N_{gal\,\mathrm{Exp}}$ & $N_{gal\,\mathrm{deV}}$   \\ \hline
      A1     & $9.69<\log(M_\ast/M_\odot)<9.89$ & $1.7<C<2.5$ & $0.04<z<0.06$ & 797 & 117  \\
      A2     & $9.69<\log(M_\ast/M_\odot)<9.89$ & $2.5<C<3.3$ & $0.04<z<0.06$ & 66 &  501  \\
      \\
      B1     & $9.89<\log(M_\ast/M_\odot)<10.09$ & $1.7<C<2.5$ & $0.05<z<0.07$ & 1028 & 175  \\
      B2     & $9.89<\log(M_\ast/M_\odot)<10.09$ & $1.7<C<2.5$ & $0.05<z<0.07$ &  83 & 1111  \\
      \\
      C1     & $10.09<\log(M_\ast/M_\odot)<10.29$ & $1.7<C<2.1$ & $0.05<z<0.08$ & 638 & 15  \\
      C2     & $10.09<\log(M_\ast/M_\odot)<10.29$ & $2.1<C<2.5$ & $0.05<z<0.08$ & 752 & 308  \\
      C3     & $10.09<\log(M_\ast/M_\odot)<10.29$ & $2.5<C<2.9$ & $0.05<z<0.08$ & 121 & 1499  \\
      C4     & $10.09<\log(M_\ast/M_\odot)<10.29$ & $2.9<C<3.3$ & $0.05<z<0.08$ & 2 & 1071  \\
      \\
      D1     & $10.29<\log(M_\ast/M_\odot)<10.49$ & $1.7<C<2.1$ & $0.05<z<0.09$ & 342 & 38 \\
      D2     & $10.29<\log(M_\ast/M_\odot)<10.49$ & $2.1<C<2.5$ & $0.05<z<0.09$ & 534 & 535 \\
      D3     & $10.29<\log(M_\ast/M_\odot)<10.49$ & $2.5<C<2.9$ & $0.05<z<0.09$ & 89  & 1468\\
      D4     & $10.29<\log(M_\ast/M_\odot)<10.49$ & $2.9<C<3.3$ & $0.05<z<0.09$ & 1 & 2153 \\
      \\
      E1     & $10.49<\log(M_\ast/M_\odot)<10.69$ & $1.7<C<2.1$ & $0.06<z<0.11$ & 239 & 74 \\
      E2     & $10.49<\log(M_\ast/M_\odot)<10.69$ & $2.1<C<2.5$ & $0.06<z<0.11$ & 555 & 1093 \\
      E3     & $10.49<\log(M_\ast/M_\odot)<10.69$ & $2.5<C<2.9$ & $0.06<z<0.11$ & 72 & 1981 \\
      E4     & $10.49<\log(M_\ast/M_\odot)<10.69$ & $2.9<C<3.3$ & $0.06<z<0.11$ & - & 1867 \\
      \\
      F1     & $10.69<\log(M_\ast/M_\odot)<11.09$ & $1.7<C<2.1$ & $0.09<z<0.13$ & 199 & 264 \\
      F2     & $10.69<\log(M_\ast/M_\odot)<11.09$ & $2.1<C<2.5$ & $0.09<z<0.13$ & 303 & 1510\\
      F3     & $10.69<\log(M_\ast/M_\odot)<11.09$ & $2.5<C<2.9$ & $0.09<z<0.13$ & 76 & 2919\\
      F4     & $10.69<\log(M_\ast/M_\odot)<11.09$ & $2.9<C<3.3$ & $0.09<z<0.13$ & 1 & 4180 \\
      \\
      G1     & $11.09<\log(M_\ast/M_\odot)<11.69$ & $1.7<C<2.1$ & $0.14<z<0.18$ & 6 & 47 \\
      G2     & $11.09<\log(M_\ast/M_\odot)<11.69$ & $2.1<C<2.5$ & $0.14<z<0.18$ & 11 & 220\\
      G3     & $11.09<\log(M_\ast/M_\odot)<11.69$ & $2.5<C<2.9$ & $0.14<z<0.18$ & 15 & 792\\
      G4     & $11.09<\log(M_\ast/M_\odot)<11.69$ & $2.9<C<3.3$ & $0.14<z<0.18$ & 3 & 2794\\

      \\

      \hline
    \end{tabular}
  \end{center}   
  \label{tbl:samples}
\end{table*}

\section{Flux Corrections to the \texttt{Model} magnitudes}
\label{sec:corrections}
In this section, we derive corrections to the original SDSS \texttt{Model} magnitudes derived from the standard
DR7 photometric pipeline (\texttt{photo\,v5\_4}). We first demonstrate that the original SDSS \texttt{Model}
magnitudes are biased, in agreement with other studies \citep{Simard, Bernardi}.  We then proceed to derive
corrections to the \texttt{Model} magnitudes.

\subsection{Systematic Biases in \texttt{Model} magnitudes}
\label{sec:pdf}
The original \texttt{Model} are affected by two sources of systematic bias
related to over-subtraction of the sky background \citep{linden2007,
Bernardi2007} and to simplistic choices of the model for the surface
brightness profile of the galaxy (exponential or de Vaucouleurs).  In this
section, we allow for more complex models to describe the light profiles
and we also allow the sky background to vary.

We fit 2D axisymmetric models (single S\'{e}rsic and double S\'{e}rsic
models along with a constant background) using the Bayesian analysis
described by \citet{DSouza} to individual postage-stamp cutouts of the
highest stellar mass and high-concentration galaxies
($11.49<\log(M_\ast/M_\odot)<11.69$, $2.9<C<3.3$ ) in the redshift range
$0.14<z<0.18$ (covered by the sample G4 above - 38 galaxies) and
$0.2<z<0.4$ (414 galaxies). The choice of the sample was motivated by the
idea of testing the robustness of the \texttt{Model} magnitudes in the
limits of high stellar mass and high redshift, where the relative
contribution due to the sky background becomes increasingly significant.

\begin{figure}
  \begin{center}
    \includegraphics[width = 0.48\textwidth]{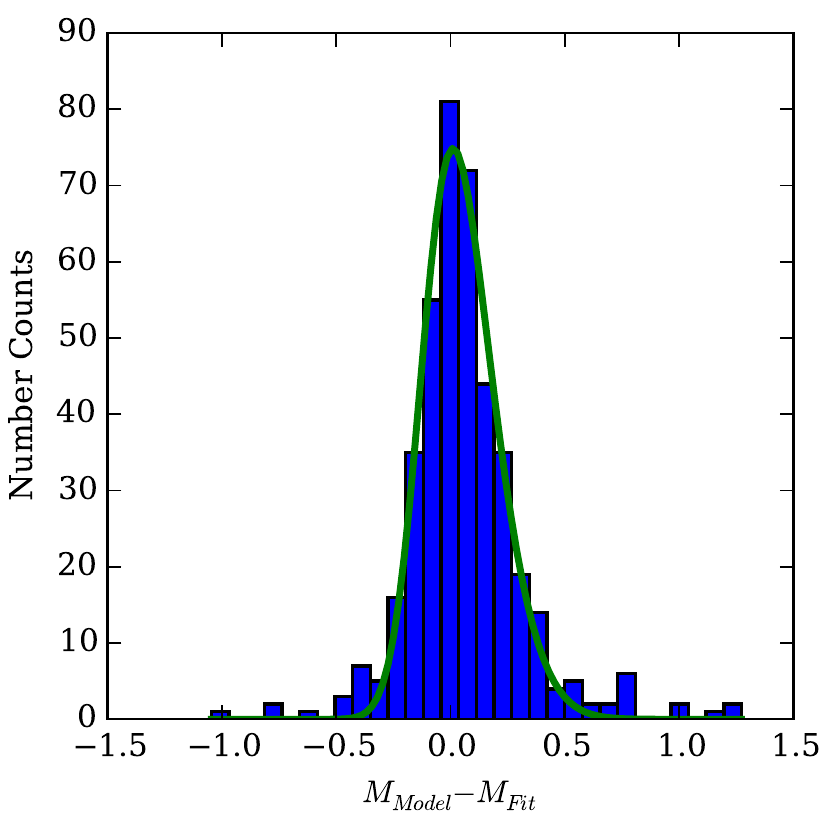}
  \end{center}
  \caption{Bias in SDSS \texttt{Model} magnitudes: A histogram of the difference between the flux derived from
    our best fit models of high stellar mass high-concentration galaxies and the \texttt{Model} magnitudes from DR7
   for galaxies with $11.49<\log(M_\ast/M_\odot)<11.69$, $2.9<C<3.3$, $0.14<z<0.18$ and $0.2<z<0.4$.
   The green solid line indicates the fit to the skewed-normal distribution.}
  \label{fig:hist_diff} 
\end{figure}

We compare our best fitting model with the \texttt{Model} magnitudes
reported by the SDSS \texttt{photo\,v5\_4} pipeline.  In Figure
\ref{fig:hist_diff}, we plot a histogram of $M_{model}-M_{fit}$  for each
galaxy. The distribution is broad with a standard deviation of 0.25
magnitudes and is positively skewed. The median is shifted by is 0.03
magnitudes and the mean by 0.08 magnitudes.  The large spread in the
histogram arises from a degeneracy between the best-fit model and the level
of sky background. The results in Figure \ref{fig:hist_diff}  demonstrate
that shallow single-exposure SDSS images are insufficient to accurately
quantify the total amount of light in massive early-type galaxies to better
than 0.25 mag. We also note that estimates of the total  flux  from a
single SDSS image will also be affected by the deblending and masking
algorithm \footnote{In this paper, we follow the deblending and masking
technique outlined in D'Souza et al. (in preparation).}.

Because of the limitations in estimating total fluxes from single SDSS
images, we have chosen to correct \texttt{Model} magnitudes using {\em
stacked images},  where the increased signal-to-noise ratio better
constrains both the model and the level of sky background.

\subsection{Stacking images}
In order to derive the flux corrections to the \texttt{Model} magnitudes, 
we used the sky-subtracted SDSS Data Release 9 images to create mosaics in the $g$, $r$ and $i$ bands 
centred on each galaxy in the sub-samples defined in Section \ref{sec:sample_stacks}. 
The mosaics extend out to radii of 0.6  - 1 Mpc depending on the stellar mass and redshift range. 
We follow the stacking procedure outlined in D'Souza et al. (in preparation) and similar to that used by \cite{DSouza}.
In short, each mosaic was deblended, masked, corrected for galactic extinction \citep{Schlegel}, transformed 
to the highest redshift in that respective bin, rotated so that the major axis of each galaxy is aligned, 
and then stacked using the truncated-mean algorithm.\footnote{For the truncated-mean stack,
we removed 5\% of the extreme minimum and maximum values for each pixel.} The $g$- and the $i$-
band images were only used to create the final mask along with the $r$-band images. 
Conservative masking was used. The final stacking was done using the masked and transformed $r$-band images.

\subsection{Measuring the Total Flux of the Stacked Images}
\label{sec:measuring_models}
Measuring the total integrated flux of a galaxy stack by fitting a model to its light distribution misses
a fair amount of light due to the inability of the model to reproduce the bulge/disk component of the inner
part of the galaxy. For example, a double S\'{e}rsic model can miss upto 0.22 mag near the centre of a galaxy
stack. On the other hand, ``isophotal'' magnitudes are unable to measure the LSB features of the galaxy stack,
especially in the low S/N regime.

In order to measure the total integrated light in each stack, we consider, therefore, a hybrid approach between
a ``model'' and an ``isophotal'' magnitude. In particular, we first fit the large mosaics using two-dimensional
axisymmetric double S\'{e}rsic models with a flat background using the Bayesian analysis described by \citet{DSouza}.
During this fit, the inner component is truncated outside a radius equal to $7\, R_{e}$, while the outer component extends out to infinity.
Then, to the flux derived by the double S\'{e}rsic model, we add the total residual flux ($=\texttt{Data}-\texttt{Model}$)
within a circular aperture of limiting radius $R_{lim}$, defined as the radius at which the residual flux is maximised.
The advantage of this hybird approach is that the double S\'{e}rsic model measures the slow decline of flux into the low S/N regime,
while the sum of the residual and model fluxes reproduces the total flux in the high S/N part of a galaxy stack.

In the next subsections, we explore the different factors that may bias our measurements of the total flux of the stack.

\subsubsection{Bias due to Inaccurate Sky Background Subtraction}
The sky residuals in the individual SDSS DR9 images are responsible for some small amount of 
residual sky background in the final stacks ($<\, 4\, \times 10^{-4}\, \mathrm{nanomaggies}$ per pixel).
We quantify this residual by adding a flat background component as a free  parameter of our models
(Section \ref{sec:measuring_models}). As show later in Section \ref{sec:models}, this bias is minimal ($\le 0.01 $ mag).

For the individual DR9 images, \cite{Blanton2011} quantified the spread in the sky background 
residuals to be $\sigma \sim 3.13 \, \times \, 10^{-3} \, \mathrm{nanomaggies}$ per pixel.
The median bias in the $r$-band magnitudes was estimated to be at the most 0.1 mag independent of $R50$. 
In addition, higher stellar mass galaxies are found predominately at higher redshifts in the SDSS spectroscopic sample,
limiting the bias in the flux of individual images caused by faulty sky-subtraction.

\subsubsection{Bias due to Models Used}
\label{sec:models}
In order to test how the choice of  model may affect the corrections to the total luminosity using
the ``hybrid'' magnitudes,  we fit each of our galaxy stack using  
two-dimensional exponential,  de Vaucouleurs,  S\'{e}rsic and  double S\'{e}rsic 
models. Each model also includes a flat sky residual. By comparing the evidences generated from the 
Bayesian fitting, we find that the double S\'{e}rsic models are preferred by more than 10-$\sigma$ 
over the other models in all cases. The de Vaucouleur model gives the highest estimate of the total 
amount of light, followed by the single S\'{e}rsic, the double S\'{e}rsic and the exponential 
model respectively. Calculating the magnitudes in the ``hybrid'' manner as described above gives very
little difference in the total flux derived from different models.

Each model also yields different estimates of the residual background in the stacks. However, determining the
background level independently and keeping it fixed during the fitting process does not alter our estimates
of the extra light (at the 0.01 mag level). This is due to the fact that the results of the fitting are driven
primarily by the inner high SNR part of the galaxy stack. 

We conclude that the combination of the depth of our stacked image and our ``hybrid'' magnitudes
enables us to accurately constrain the total flux in the galaxy stack. Our outer models are not truncated, 
but instead extend out to infinity. The difference between models which are truncated at $7\, R_{e}$ and 
models which instead extend out to infinity is at most 0.05 mag.

\subsection{Measuring the Flux Corrections}
\begin{figure*}
  \begin{center}
    \includegraphics[width = \textwidth]{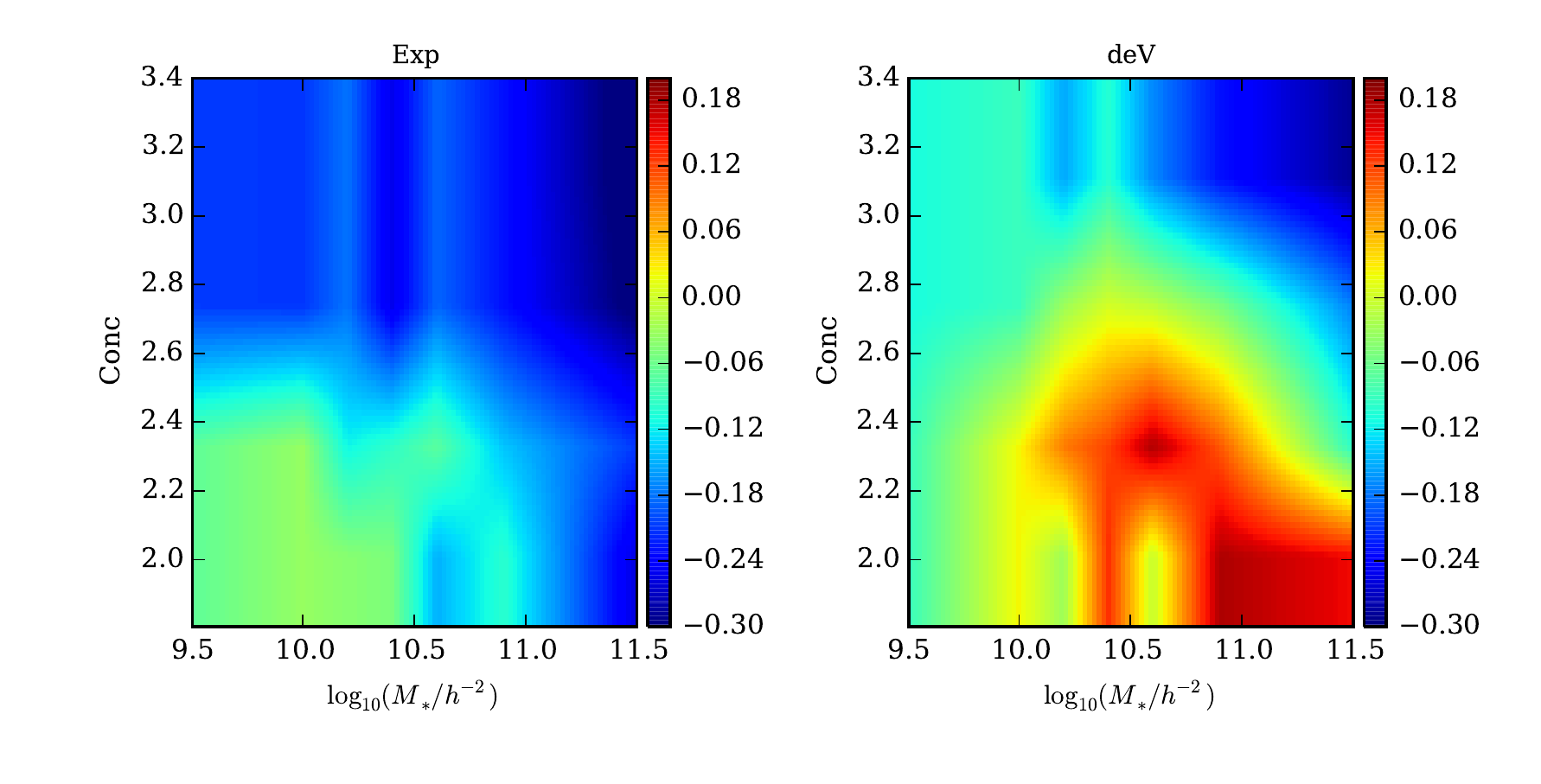}
  \end{center}
  \caption{The flux corrections ($\Delta \mathrm{Mag}$) as a function of stellar mass and concentration using an
    interpolation scheme for  exponential (\texttt{Exp}) and de Vaucouleurs (\texttt{DeV}) fit galaxies.}
  \label{fig:extra_light} 
\end{figure*}

For each stellar mass, concentration range and model fit type (exponential or  de Vaucouleurs),
we measure the average extra flux correction to the \texttt{Model} magnitudes as the difference between the total integrated light
in the stack and the median \texttt{Model} flux of the galaxies in the stack. The median \texttt{Model}
magnitude was calculated by taking the median of the individual fluxes of galaxies in the stack. We find that
the median \texttt{Model} magnitude is on average higher than the mean \texttt{Model} magnitudes.
We use a two-dimensional-interpolation scheme to calculate the average extra light as a continuous function
of stellar mass and concentration for each model type. These are shown in Figure \ref{fig:extra_light}.
As can be seen, there is an extra light contribution from those galaxies which were fit by an exponential model 
both for high concentrations and for high stellar masses. The extra light correction from those galaxies fit by a de
Vaucouleurs model comes predominately from the massive, high concentration galaxies. On the other hand, the de Vaucouleurs
model often over-estimates the flux of a galaxy for low concentration massive galaxies.

\begin{figure}
  \begin{center}
    \includegraphics[width = 0.48 \textwidth]{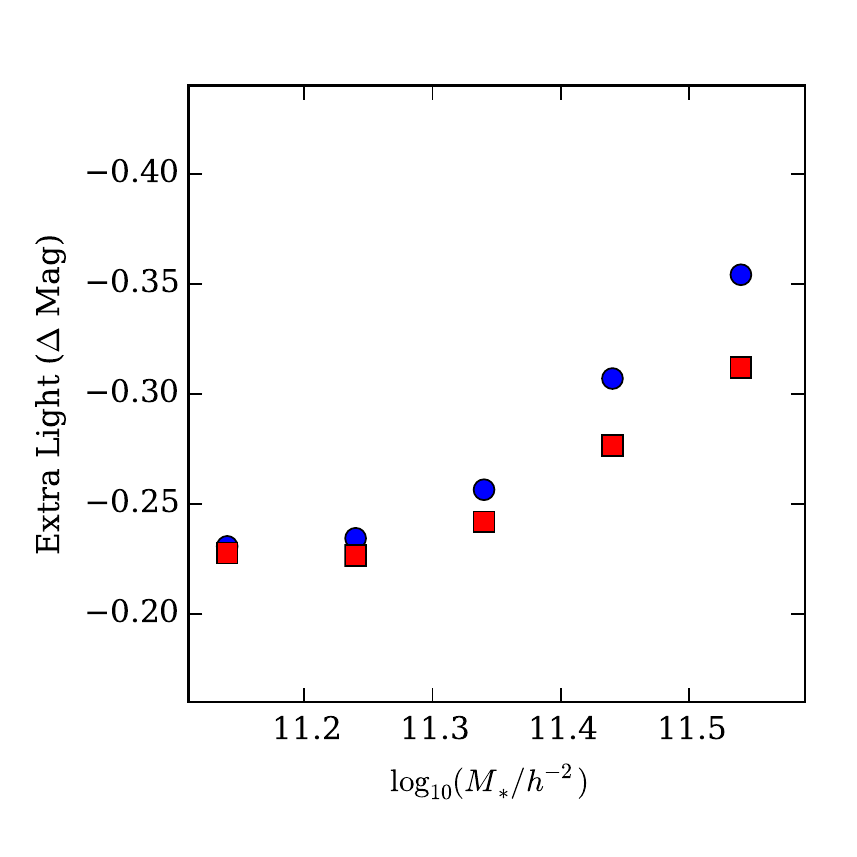}
  \end{center}
  \caption{The flux corrections ($\Delta \mathrm{Mag}$) as a function of stellar mass for galaxies in the sample G4 (blue circles).
    We also indicate the uncertainty in the corrections by showing the flux corrections derived from the mean \texttt{Model}
    flux of the stacks (red circles).}
  \label{fig:extra_light2} 
\end{figure}

We note that the large width of the stellar mass bins for the highest stellar mass galaxies may influence  the correction
derived in the stacking procedure. To account for this,  we divide our highest mass sample, G4 ($11.09<\log(M_\ast/M_\odot)<11.69$, $2.9<C<3.3$,
$0.14<z<0.18$ ), into smaller mass bins of size 0.1 dex. We find that the relative corrections ranges from 0.23 to 0.31 mag, gradually
increasing from the lowest to the highest stellar mass bin (see Figure \ref{fig:extra_light2}). The mean correction derived by stacking
the entire sample G4 is 0.29 mag.

For galaxies outside the the mass limits defined in Section \ref{sec:sample_stacks}, 
we extrapolate assuming the same mass corrections of the nearest defined mass bin.
In particular, at the high mass end, there are 116 galaxies with stellar masses larger than 
$\log(M_\ast/M_\odot)>11.69$, the highest stellar mass bin used above. For these galaxies, we assume 
the corrections to be the same as found for the highest stellar mass bin (~0.31 mag).

Assuming a constant M/L for each galaxy, we calculate the extra mass for each galaxy in our main sample given its
stellar mass, concentration and model type (by comparing the likelihoods of the \texttt{Model} fits 
from the SDSS database) as:
\begin{equation}
  \log \frac{M_{\ast} + \delta M_{\ast}}{M_{\ast}} = - \Delta \mathrm{Mag}/2.5
\end{equation}

\section{The Stellar Mass Function of Galaxies}
\label{sec:massfunction}

\subsection{Method}
We estimate the abundance of galaxies as a function of their stellar mass using  the $1/V_{max}$ method outlined
by \cite{Li2009}. In combination with the depth and the large spectroscopic sample of SDSS, the $1/V_{max}$
method provides an unbiased estimate of the stellar mass function and its normalisation.In Section \ref{subsec:variance},
  we demonstrate that the $1/V_{max}$ estimator is unbiased against large scale structure at stellar masses of $\log(M_\ast/M_\odot) \ge 9.5$,
  which is the regime studied in this work. We limit ourselves to this regime since as estimated by Figure 4 of \cite{Baldry08},
all galaxies above stellar masses of $\log(M_\ast/M_\odot) \ge 9.5$, will be detected irrespective of their central
surface brightness. Moreover, our flux corrections begin from $\log(M_\ast/M_\odot) \ge 9.6$ upwards.

For each observed galaxy $i$, we define the quantity $z_{max,i}$ to be the maximum redshift at
which the observed galaxy would satisfy the apparent  magnitude limit of our sample
$m_{r_{Pet}}\,\le\,17.6$. Evolutionary and K-corrections are included when  calculating $z_{max,i}$.
Hence, $z_{max,i}$ is the minimum of the upper limit of the redshift slice and the solution
of the equation:
\begin{equation}
  \label{eq:zmax}
  M_{i} = m^{Faint}_{r_{Pet}} - DM(z_{max}) - K(z_{max}) + Q_{e}(z_{max}-z_{i})
\end{equation}

Similarly, we also define $z_{min,i}$ as the minimum redshift at which the
galaxy would be present in our sample. Hence,  $z_{min,i}$ is the maximum of the lower limit
of the redshift slice and the solution to the equation:
\begin{equation}
  \label{eq:zmin}
  M_{i} = m^{Bright}_{r_{Pet}} - DM(z_{min}) - K(z_{min}) + Q_{e}(z_{min}-z_{i})
\end{equation}

This then allows us to calculate $V_{max,i}$ for the galaxy
in question as the total co-moving volume of the survey between $z_{min,i}$ and  $z_{max,i}$. 
The stellar mass  function can be then estimated as:
\begin{equation}
  \Psi(M_{\ast}) \Delta M_{\ast} = \sum_{i} (f_{norm\,coll,i}\, V_{max,i})^{-1} \,
\end{equation}
where  $f_{norm\, coll,i}$ is the normalised fiber collision
factor defined below, and the sum extends over all sample galaxies with
stellar masses in the range $M_{\ast} \pm 0.5 \,\delta M_{\ast}$. The error bars are 
estimated by taking into consideration both Poissonian and bootstrapping errors,
as well as errors due to cosmic variance (See \ref{subsec:variance}).

We calculate the stellar mass function in the total redshift range $0.001 \le z \le 0.5$ 
as well as in three redshift slices: $0.001 \le z \le 0.15$, $0.15 \le z \le 0.3$ and $0.3 \le z \le 0.5$.

\subsection{Robustness of the $1/V_{max}$ Estimator}
\label{subsec:variance}
\begin{figure}
  \begin{center}
    \includegraphics[width = 0.48\textwidth]{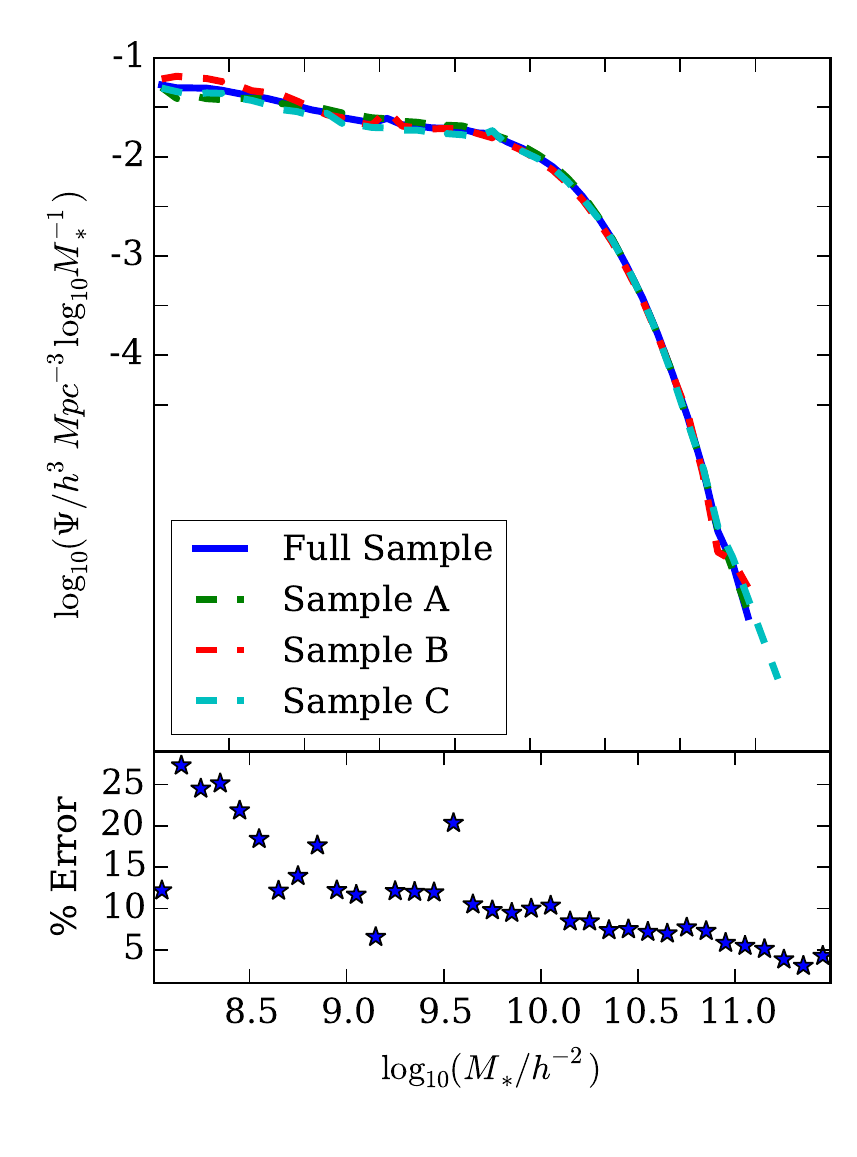}
  \end{center}
  \caption{Top panel: The stellar mass function for the complete sample as well as the three independent smaller samples A, B and C,
    using the MPA-JHU stellar masses for the full redshift range $0.001 \le z \le 0.5$. 
    Bottom Panel: The standard deviation in the stellar mass function as a function of stellar mass.
  }
  \label{fig:variance} 
\end{figure}

In this work, we estimate the abundance of galaxies using the $1/V_{max}$ method. Given the large effective surface area (nearly ~$6570\,deg^{2}$)
and the depth of spectroscopic sample, the $1/V_{max}$ method will be invariant to large-scale structure up to a limiting stellar mass. To test this,
we divide our sample into three independent but contiguous parts (Sample A, Sample B and Sample C split by right ascension), and calculate the standard
deviation in the stellar mass function as a function of stellar mass. In the bottom panel of Figure \ref{fig:variance}, the standard deviation is plotted as
a function of stellar mass. The difference in the estimates of the stellar mass function due to the  $1/V_{max}$ method from the three independent
samples is less than 10\% for stellar masses $log(M_\ast/M_\odot) > 9.5$. The standard deviation gives us also a handle on the errors in our estimates
of the stellar mass function due to the cosmic variance.

\subsection{The Effect of Systematic and Random Errors on the SMF} 

In calculating the stellar mass function, various systematic and random effects combine to affect the final
result. We discuss each of these effects in turn in the following subsections:

\subsubsection{MPA-JHU Stellar Masses and Extra light from Photometry}
The first source of systematic bias comes from the estimation of the stellar mass of individual galaxies.
In this work, we use the MPA-JHU stellar masses to calculate the stellar mass function.
This involves a change of flux (from \texttt{Petrosian} to \texttt{Model} magnitudes)
and M/L ratio (from NYU-VAGC to MPA-JHU) relative to \cite{Li2009}.

We find that the use of NYU-VAGC stellar masses based on the \texttt{Model} magnitudes rather
than the \texttt{Petrosian} magnitudes introduces a shift beyond the knee of the
stellar mass function towards a shallower slope at the higher mass end. This shift is then
further increased when we switch to MPA-JHU stellar masses based on the \texttt{Model}
magnitudes. The slope of the massive end of the mass function is shallower than that
obtained by shifting the mass function derived from  the NYU-VAGC stellar masses by 0.1 dex (See appendix of
\citealt{Li2009}: $\Delta \log M_{*} = 0.1$). At a stellar mass of $\log M_{*} \sim 11.5\,\msun$,
this accounts for an increase in the stellar mass function by a total of 1.24 dex (a 0.57 dex increase due to
the change from \texttt{Petrosian} to \texttt{Model} magnitudes and a 0.67 dex increase due to the
change from the  NYU-VAGC to MPA-JHU M/L ratios).

Our assumed M/L ratio affects our estimation of the stellar mass function. The use of the MPA-JHU
stellar M/L ratios  makes the slope at the massive end shallower than the NYU-VAGC M/L ratios.
We note that the MPA-JHU M/L ratios are derived from models that include the possibility of
complex star formation histories, whereas the NYU-VAGC assumes that red galaxies can be described
by single stellar populations. Analysis of spectra of massive galaxies
in the BOSS survey by \cite{Chen2013} indicates that the star formation histories of the most massive
galaxies are characterised by episodic star formation histories.

The extra flux derived from the photometry of stacked galaxies introduces a further shift, making the slope 
at the massive end of  the stellar mass function even shallower. We find that this shift of the stellar mass
function is independent of whether we apply the corrections only to the central galaxies,
or to all the galaxies in the sample. Although a small difference is found at the knee of the mass function,
both results are consistent with each other within the error bars.

\subsubsection{Fiber Collisions}
The second source of systematic bias is caused by fiber collisions.
The NYU-VAGC catalogue lists the spectroscopic completeness $f_{sp}$ of each galaxy, 
defined as the fraction of photometrically defined target galaxies in the subarea 
for which usable spectra are obtained. The NYU-VAGC catalogue calculates the average 
completeness for each of these subareas by taking into consideration overlapping plates. 
In the jargon of the NYU-VAGC catalogue, these subareas are called \emph{sectors}. $f_{sp}$
contains information about the missing galaxies due to lack of fibers in dense regions, missing
galaxies due to spectroscopic failures, and missing galaxies due to fiber  collisions. The average $f_{sp}$ for the sample defined above is 0.9146.
However, $f_{sp}$ assumes that all galaxies with  measured spectra are randomly distributed
within a sector, and hence cannot account for specific differences between high and low density 
regions in the same sector. In particular, due to fiber collisions, certain galaxies (e.g. satellite galaxies of 
large clusters found at high redshifts) will be preferentially missed. 

To account for fiber collisions, we define the fiber collision $f_{coll,i}$ for each galaxy, 
as the fraction of  photometrically defined target galaxies that fall within a area of 55'' in 
radius. $f_{coll,i}$ takes the values between 0.111 and 1.0 (that is, 8 closest 
neighbours and no neighbours respectively). The average of $f_{coll,i}$ over our whole sample 
is 0.93819. We normalise $f_{coll,i}$ such that it's average value is the same as that of 
$f_{sp}$. $f_{norm \,coll,i} = fac* f_{coll,i}$, where $fac$ is defined as $<f_{sp}>/<f_{coll}>$ 
and takes the value 0.9749. The normalised fiber collision $f_{norm \, coll}$ now has the general
average properties of $f_{sp}$, but can better account for fiber collisions.

Weighting by normalised fiber collision maintains the normalisation of the stellar mass function 
at the low mass end and increases the mass function up to 22\% at the high mass end.
\cite{Li2009} did not include fiber collisions in their derivation and we can
only reproduce their stellar mass function by using \texttt{Petrosian} magnitudes
and by neglecting the effect of $f_{sp}$. 

\subsubsection{Evolution Corrections}
The third main source of systematic error is related to the assumption about the passive evolution of galaxies both in their
number density and luminosity. In order to construct a stellar mass function from a large redshift range (($0.001 \le z \le 0.5$),
we would need account for the passive evolution of galaxies using a so-called evolutionary correction. Assuming such a uniform
evolutionary correction is problematic, since galaxy evolution is a function of galaxy type and cannot be described by a simple linear model. 
For example, star-forming galaxies will evolve more slowly in luminosity than early-type galaxies.

In order to quantify the effects on the stellar mass function related to the assumptions about galaxy evolution, we
consider two approaches. In the first approach, we assume a uniform  evolutionary correction ($Q_{r}=1.62$), which would
represent an upper limit for the evolution of early-type galaxies with high stellar masses and stellar populations that evolve passively with time (i.e. in the 
absence of any mergers). In the second approach, we derive the stellar mass function without evolution in 
three redshift slices:  $0.001 \le z \le 0.15$, $0.15 \le z \le 0.3$ and $0.3 \le z \le 0.5$.

\begin{figure}
  \begin{center}
    \includegraphics[width = 0.48\textwidth]{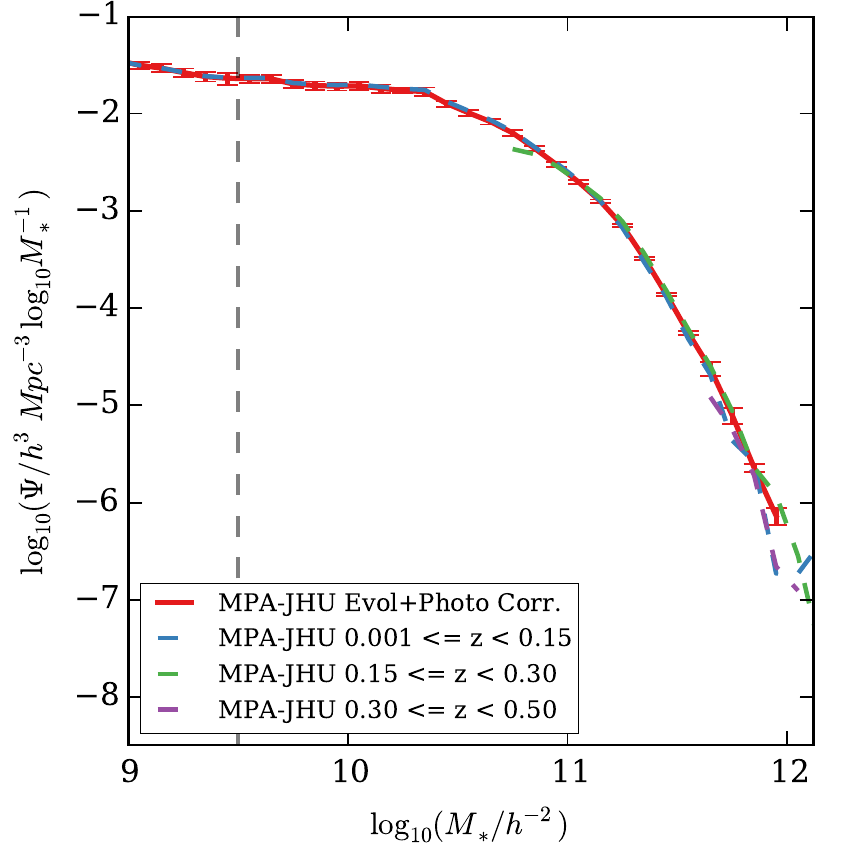}
  \end{center}
    \caption{
      The effect of evolutionary corrections on the stellar mass function: The red solid line shows the stellar mass function
      calculated using MPA-JHU stellar masses corrected for missing light from photometry of stacked
      galaxies, corrected for fiber collisions and with uniform evolutionary correction $Q=1.62$ in the redshift range $0.001 \le z \le 0.5$.
      Also plotted are the stellar mass function without evolutionary corrections in redshift slices
      $0.001 \le z \le 0.15$, $0.15 \le z \le 0.3$ and $0.3 \le z \le 0.5$ dashed blue, green and violet lines.
        }
  \label{fig:massfunction_evolution} 
\end{figure}

In Figure \ref{fig:massfunction_evolution}, we plot the stellar mass function derived using the MPA-JHU
stellar masses, including a uniform evolutionary correction, accounting for fiber collisions and for the additional
stellar mass corrections due to the extra light at large radii (red solid curve). 
In addition, we also indicate the mass function calculated in the three redshift slices mentioned above, without evolution.
As seen from Figure \ref{fig:massfunction_evolution}, the evolutionary correction has only a small effect on the stellar mass function ($\sim 10\%$ at the massive end).
This is related to the fact that the luminosity evolution is implicitly folded into the derivation of the M/L ratio.

\subsubsection{Uncertainty due to binning the data}
Another source of systematic bias is related to binning the data in calculating the mass function via the $1/V_{max}$ method. 
In particular, this introduces further uncertainty at the massive end of the mass function due to a combination of the low number
statistics and the steep slope of the mass function 
over this mass range. In order to quantify this uncertainty, we recalculate 
the mass function with different values for the bin sizes, from 0.05 dex to 0.4 dex. 
In particular, larger bin sizes tends to bias the slope at the high mass end of 
the mass function towards shallower values. Reducing the bin size increases the 
steepness of the slope until a saturation limit of about 0.1 dex. The variation 
caused by changes in the bin size around the saturation limit is within the uncertainties 
derived by bootstrapping and within the Poissonian errors. Hence, we calculate the 
stellar mass function in bins of 0.1 dex.

\subsubsection{Eddington Bias}
Another source of systematic bias in the stellar mass function is caused by the random errors in the flux and 
M/L ratios of individual galaxies. Such an ``Eddington'' bias causes the stellar mass function
to be higher in the low-number density part because of scattering from the lower stellar masses (higher number density).
This becomes particularly acute because of the steepness of the stellar mass function at higher stellar masses.

To correct for this bias, we assume a parametrized form for the stellar mass function. We convolve this function with a distribution of
the uncertainties in the stellar mass. We then fit this convolved function to the binned values of the stellar mass function calculated from
the data using a maximum-likelihood method. The best fit paramteric function is thus our true stellar mass function corrected
for the Eddington bias. For the parametric function, we assume a double Schechter function, given by
\begin{eqnarray}
  \label{eq:schechter}
  \nonumber
  \Psi_{M} \mathrm{d} M & = &  \bigg[\, \frac{\Psi^{\ast}_{1}}{M^{\ast}_{1}}\,e^{-M/M^{\ast}_{1}} \left(\frac{M}{M^{\ast}_{1}} \right)^{\alpha_{1}}  + \\
         & &    \,\,  \frac{\Psi^{\ast}_{2}}{M^{\ast}_{2} } \,e^{-M/M^{\ast}_{2}} \left(\frac{M}{M^{\ast}_{2}} \right)^{\alpha_{2}}  \bigg] \mathrm{d} M  \, ,
\end{eqnarray}
where $\Psi_{M} \mathrm{d} M$ is the number density of galaxies between
$M$ and $M + \mathrm{d} M$. This provides a much better fit to the data relative to a single Schechter.  We further assume that the uncertainties in the stellar mass are distributed normally in $\log_{10} (M_{\ast}/M_\odot)$.

To estimate the uncertainties in the stellar mass, we first estimate the M/L uncertainties as a function of stellar mass 
from the MPA-JHU database. We find that the average uncertainty $\Delta \log_{10} (M/L)$ ranges from 0.08 to 0.1 as a function of
stellar mass. We then estimate the average uncertainty in the \texttt{Model} magnitude as a function of stellar mass. We find that the
average uncertainty in the \texttt{Model} magnitude is $\sim\,0.02\,mag$ across the stellar mass range considered.
Hence the M/L uncertainty is much larger than the flux uncertainty.
  
We find that correcting the stellar mass function for the Eddington bias reduces it at the high mass end by as much as 0.48 dex.

\subsection{Results: Stellar Mass Function}
\label{sec:fits}
\begin{figure}
  \begin{center}
    \includegraphics[width = 0.48\textwidth]{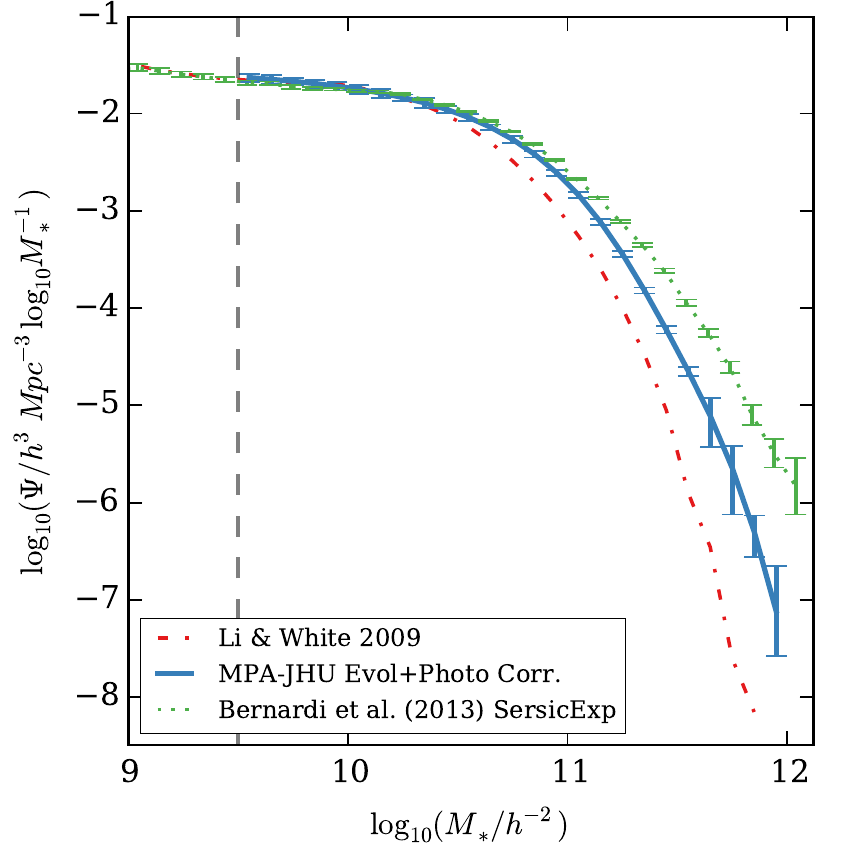}
  \end{center}
    \caption{
      The stellar mass function: The blue solid line shows the stellar mass function
      calculated using MPA-JHU stellar masses, corrected for missing flux, fiber collisions, evolution and Eddington bias.
      The red dot-dashed line shows the original Li \& White 2009 stellar mass function calculated using the NYU-VAGC stellar
      masses based on \texttt{Petrosian} magnitudes. Also shown are the Bernardi et al. (2013) stellar mass function values (green) based on 
      Sersic-Exp fits to individual galaxies. The dashed vertical line indicates our lowest stellar mass limit
      above which the stellar mass function is not affected by surface brightness completness issues.
    }
  \label{fig:massfunction1} 
\end{figure}

In Figure \ref{fig:massfunction1}, we present our final estimate of the stellar mass function corrected for 
missing flux, fiber collisions, evolution and Eddington bias with that of the original Li \& White (2009) in red
and the Bernardi et al. (2013) (Sersic-Exp fits) stellar mass function in green.

We provide a parametric representation of the stellar mass function for stellar masses greater than
$\log(M_\ast/M_\odot) \ge 9.5$. The parameters of the double  Schechter function are listed in Table \ref{tbl:schechter}.
An integration of our stellar mass function for stellar masses greater than $\log(M_\ast/M_\odot) \ge 9.5$
gives the mean comoving stellar mass density of the low redshift universe as $\phi_{*}=3.7 \pm 0.3 \,10^{8}\,h$Mpc$^{-3}$.
This amounts to a 35\% increase in the mean comoving stellar mass density contributed from the same stellar mass range 
for the \cite{Li2009} stellar mass function.
In particular, focussing on the high stellar mass end: the mean comoving stellar mass density of galaxies with stellar masses
$\log(M_\ast/M_\odot) \ge 11.0$ is a factor of 3.36 larger than the estimate by Li \& White (2009), but is 43\% smaller than
reported by Bernardi et al. (2013).

\begin{table}
\caption{Parameters of a double Schechter function fit to the stellar
  mass function of SDSS galaxies. 
}
\begin{center}
\begin{tabular}{ccr} \hline\hline

\multicolumn{1}{c}{$\Phi^{\ast}$}  & \multicolumn{1}{c}{$\alpha$}  &  \multicolumn{1}{c}{$\log_{10}M^\ast$}   \\
\multicolumn{1}{c}{($h^{3}$Mpc$^{-3}\log_{10}M^{-1}$)}                & \multicolumn{1}{c}{}      &      \multicolumn{1}{c}{($h^{-2}M_\odot$)}\\  \\  \hline
0.008579 &  -1.082  & 10.615  \\
0.000355 &  -1.120  & 10.995\\ \hline
\end{tabular}
\end{center}
\label{tbl:schechter}
\end{table}

\section{Galaxy Luminosity Function}
\label{sec:lumfunction}
\begin{figure}
  \begin{center}
    \includegraphics[width = 0.48\textwidth]{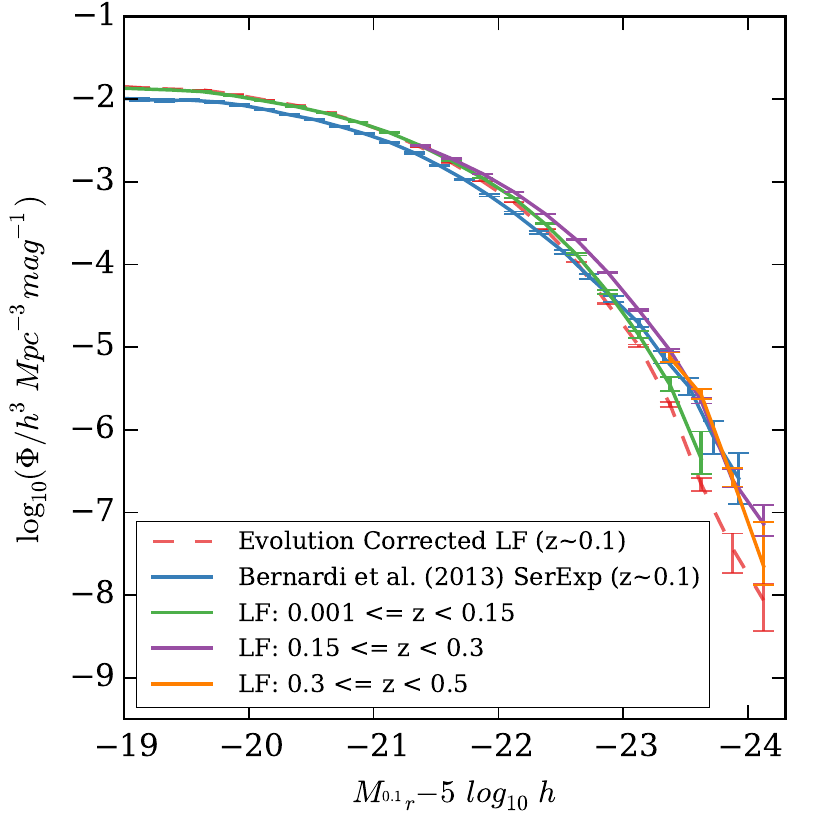}
  \end{center}
  \caption{The luminosity function: The $M_{{}^{0.1}r}$ luminosity function calculated
    with photometric corrections, fiber collisions and flux uncertainty in three redshift slices and assuming an uniform evolutionary correction of  $Q_{r}=1.62$}.
    We also show the corresponding luminosity function from Bernardi et al. (2013) Sersic-exponential fits. 
  \label{fig:luminosityfunction} 
\end{figure}

Similar to the galaxy stellar mass function, we also calculate the galaxy luminosity function using
the $1/V_{max}$ method. However, more careful attention needs to be paid to the evolutionary corrections
which affects the luminosity function not only via the derivation of $V_{max}$, but also via the calculation
of a galaxy luminosity via equation \ref{eq:mag}. We calculate the luminosity function using two approaches:
in redshift slices  ($0.001 \le z \le 0.15$, $0.15 \le z \le 0.3$ and $0.3 \le z \le 0.5$) without evolution
and using a uniform evolutionary correction of $Q_{r}=1.62$. In Figure \ref{fig:luminosityfunction}, we present
the results of $M_{0.1\,r}$ band luminosity function considering \texttt{Model} magnitudes with photometric
corrections from stacking, fiber collisions and evolutionary corrections in bins of 0.25 dex. We also indicate
the luminosity funtion without evolution corrections in three redshift slices. A comparision of our results with
those of \cite{Bernardi} would require a more careful treatment of luminosity evolution which is beyond the scope
of this paper.

\section{Summary}
\label{sec:summary}
In this paper, we have shown that stacking similar galaxies together in volume-limited
stellar mass and concentration bins allows one to derive average flux corrections to the
SDSS \texttt{Model} magnitudes. In particular, we find that these corrections range from
0.02 to 0.31 magnitude, depending on the stellar mass and concentration of the galaxy.

We apply these corrections to the \texttt{Model} fluxes and re-derive the stellar mass
function using MPA-JHU stellar masses, accounting for galaxy evolution corrections and 
fiber collisions. We find that the slope of the massive end of the stellar mass function 
is shallower than reported by \cite{Li2009}, but much steeper than derived by \cite{Bernardi}.

The biggest change in the slope at the massive end of the mass function comes from our adoption
of the MPA-JHU stellar masses (as much as a 1.24 dex increase at $\log M_{*} \sim 11.5\,\msun$ with
respect to \citealt{Li2009}).
This involves an increase of 0.57 dex and 0.67 dex due to the changes in flux and M/L ratio respectively. The second major contributor is the bias caused by the uncertainty in M/L
ratio and flux measurements  of individual galaxies which accounts for a decrease of $\sim\,0.48$ dex
in the mass function at the massive end. Fiber collisions contributes to an increase of nearly 22\%
at the massive end. Galaxy evolution corrections accounts for a decrease of maximum 10\% at the massive end of the mass function.

We also derive the $r$-band galaxy luminosity function and obtain similar results.
In particular, the biggest source of systematic uncertainty in the galaxy luminosity
function is related to the model assumed for the galaxy evolution correction. In this Paper, 
we use the evolution correction values derived by \cite{Blanton2003}, which serves as an upper
limit for galaxies at the bright end of the galaxy luminosity function. 

\section{Discussion}
\label{sec:discussion}
The flux corrections to the SDSS \texttt{Model} magnitude and their respective uncertainties
derived in this work by stacking mosaics of similar galaxies in volume limited stellar mass
and concentration bins are consistent with those presented by \cite{Simard}. We find no evidence
for the need of large flux corrections of the order of 0.5 magnitudes as proposed by \cite{Bernardi}.

Our results are also consistent with extremely deep imaging of nearby early-type galaxies, 
obtained with the MegaCam camera on the Canada-France-Hawaii Telescope which indicate 
that outer LSB light  contributes 5 to 16 percent to a galaxy's total luminosity \citep{duc}. 
Stacking results for luminous red galaxies (average redshift of $z \sim 0.34$) from \cite{Tal}
also indicate that typical SDSS-depth images miss about 20 percent of the total stellar light.

A number of systematic differences could contribute to the discrepancy between our results
and those by \cite{Bernardi}. In the limit of low SNR, the determination of the sky background 
level can influence the measured flux of a galaxy derived from fitting models to the 
surface brightness distribution. The depth of an image limits ones ability to distinguish 
between the flux of the outer LSB features of the galaxy and the sky background, especially 
for large stellar mass galaxies at higher redshifts. The use of multi-component 
models aggravates this problem.

The simultaneous estimation of the model parameters and the sky background level may
be prone to systematic bias, since these are often degenerate with each other.
\cite{Bernardi} use the \texttt{PyMorph} algorithm (based on \texttt{GALFIT}), which estimates
the galaxy flux based on model fitting along with a simultaneous estimation of the sky
background. \cite{Meert2013} and \cite{Meert2015} have already highlighted the effect of a
bias in the sky subtraction on the total flux of a galaxy. On the other hand, SDSS \texttt{Photo}
pipeline estimates the \texttt{Model} magnitudes by first independently estimating and
subtracting the local sky background. A similar procedure is followed by \cite{Simard}.
In this work, we use the background subtracted images provided with SDSS DR9 to derive the
flux corrections. In addition, the depth of our stacked images allows us to accurately 
determine the residual sky background.

Estimating the total flux of a galaxy is dependent on the exact procedure used
for deblending and masking ( see \citealt{Blanton2011} and \citealt{Simard}). In particular, 
the amount of masking employed has a substantial effect on the amount of flux that is 
derived for a specific galaxy. In this Paper, we use the conservative masking described 
by \cite{DSouza2}, which involves using multiple runs of \texttt{SExtractor} \citep{Bertins1}. 

\cite{Guo} calculated the stellar mass function using the NYU-VAGC stellar M/L ratios and \texttt{Model}
magnitudes using the methodology of \cite{Li2009}. The stellar mass function derived here has a large shift
and shallower slope than \cite{Guo}, owing primarily to the use of the MPA-JHU stellar masses and the
flux corrections to the \texttt{Model} magnitudes.  The results of our work will affect the majority of
recent halo occupation and abundance matching studies (e.g. Moster et al. 2013) that use the measurements
of the stellar mass function from \cite{Guo}. 

Finally, we comment that the majority of studies of the evolution of the massive end of the stellar mass
function have found suprisingly little change out to $z \sim 1$ (Maraston et al 2013; Moustakas et al 2013;
Fritz et al 2014). The co-moving number density of galaxies with stellar masses greater than
$10^{11} M_{\odot}$ has apparently remained constant over the past 9 Gyr, calling into question the
late build-up of these systems through mergers and accretion. Our work has shown that a significant fraction
of the mass of these systems may be ``hiding'' in low surface brightness outer components that
are systematically missed by conventional photometric extraction software. Accurately quantifying the
{\em evolution} of the stellar mass in these halos will be an important challenge for next
generation deep imaging surveys.

\label{lastpage}

\section*{Acknowledgements}
Funding for SDSS-III has been provided by the Alfred P. Sloan Foundation, the
Participating Institutions, the National Science Foundation, and the U.S.
Department of Energy Office of Science. The SDSS-III web site is
\url{http://www.sdss3.org/}.

SDSS-III is managed by the Astrophysical Research Consortium for the
Participating Institutions of the SDSS-III Collaboration including the
University of Arizona, the Brazilian Participation Group, Brookhaven National
Laboratory, University of Cambridge, Carnegie Mellon University, University of
Florida, the French Participation Group, the German Participation Group,
Harvard University, the Instituto de Astrofisica de Canarias, the Michigan
State/Notre Dame/JINA Participation Group, Johns Hopkins University, Lawrence
Berkeley National Laboratory, Max Planck Institute for Astrophysics, Max Planck
Institute for Extraterrestrial Physics, New Mexico State University, New York
University, Ohio State University, Pennsylvania State University, University of
Portsmouth, Princeton University, the Spanish Participation Group, University
of Tokyo, University of Utah, Vanderbilt University, University of Virginia,
University of Washington, and Yale University.

\end{document}